\documentstyle[aps,twocolumn]{revtex}
\begin{document}
\draft
\title{Theory of Shubnikov-de Haas oscillations around the $\nu =1/2$ filling
factor of the Landau level:
Effect of gauge-field fluctuations.}
\author{ A.G. Aronov $\sp{\dagger,1,2,3}$, E. Altshuler$\sp{2}$,
A.D. Mirlin$\sp{1,4}$ and  P. W\"{o}lfle$\sp{1}$}
\address{
$\sp{1}$ Institut f\"{u}r Theorie der Kondensierten Materie,
  Universit\"{a}t Karlsruhe, 76128 Karlsruhe, Germany}
\address{ $\sp{2}$ Department of Condensed Matter Physics,
The Weizmann Institute of Science, 76100 Rehovot, Israel
}
\address{
$^3$ A.F. Ioffe Physicotechnical Institute, 194021 St.Petersburg,
Russia.}
\address{
$^4$ Petersburg Nuclear Physics Institute, 188350 Gatchina, St.Petersburg,
Russia.}
\date{December 15, 1994}
\maketitle
\tighten
\begin{abstract}
We present a theory of magnetooscillations
around the $\nu =1/2$ Landau level filling factor based on a model
with a fluctuating Chern--Simons field.
The quasiclassical treatment of the problem is appropriate
and leads to an unconventional
$\exp\left[-(\pi/\omega\sb{c}\tau\sp{*}\sb{1/2})\sp{4}\right]$
behavior of the amplitude of oscillations.
This result is in good qualitative agreement with available
experimental data.
\end{abstract}
\narrowtext

Since the discovery of the Fractional Quantum Hall effect
(FQHE)\cite{tsui1}, the physical properties of a high mobility
electron gas subjected to a strong magnetic field are attracting
great interest.
Laughlin's theory \cite{laugh} gives a very good description
of the properties of the FQHE states with filling factors
$\nu=1/(2m+1)$.
The subsequently proposed hierarchy scheme \cite{hal1}
explains, in principle, the existence of FQHE states with arbitrary
filling factors $\nu=p/q$ ( $q$ is odd).
However, some drawbacks of this scheme were discovered later.
In particular, the FQHE state with $\nu=p/(2p\pm 1)$ appears
only on the $p$-th level of hierarchy, so that the scheme does not
explain why these states are experimentally dominating.
This discrepancy motivated Jain \cite{jain} to propose
a different concept based on converting the electrons
into composite fermions by attaching to them
an even number of flux quanta.
Following a similar approach Halperin, Lee and Read \cite{hlr}
developed a theory for the half filled Landau level (see also
\cite{kz}).

This theory gives an explanation for many experimentally
observed properties of the $\nu=1/2$ state,
such as a non-zero value of the longitudinal resistivity,
and an anomaly in the surface acoustic wave propagation \cite{will}.
It predicts the formation, at half filling, of a metallic state with
well defined Fermi surface.
This prediction received additional confirmation in recent
experiments \cite{kan} where a dimensional resonance of the
composite fermions was found. From this point of view,
the $\nu=p/(2p\pm 1)$ series can be considered
as the usual  $\nu=p$ Shubnikov--de Haas oscillations (SDHO) for the composite
fermions,
providing an explanation for the prominence of the above
FQHE states.
Indeed, the oscillating behavior of the longitudinal resistivity
$\rho\sb{xx}$ near $\nu=1/2$ is very much reminiscent of its behavior
in low magnetic fields where conventional SDHO take place \cite{st2,le}.

Comparison of the resistivity oscillations near $\nu=1/2$ and in a
weak  magnetic field shows, however, not only a similarity
in shape, but also an important difference: the amplitude of the
SDHO near half filling decreases on approaching  $\nu=1/2$
($p=\infty$) much faster than it does in  weak fields
on approaching zero magnetic field $B=0$ ($\nu=\infty$).
As a result, the $\nu=p/(2p\pm 1)$ oscillations vanish
at $p=p\sb{m\!a\!x}=6\div 9$ \cite{st2,le}.
This reflects a difference in the physical properties between the states
at $\nu=1/2$ and  $B=0$, calling thus for a detailed theoretical
consideration.

The crucial feature which distinguishes these states is the presence
of fluctuations of the fictitious, Chern--Simons (CS), magnetic  field
\cite{hlr}.
The scattering of the fermions by these fluctuations turns out to
be the dominating mechanism of scattering.
On the other hand, the problem of a quantum particle
in a random magnetic field in 2D was studied in our recent papers
\cite{2,1}.
We have shown that whereas the transport relaxation time $\tau\sb{t}$
can be found in the usual way within perturbation theory,
the single particle properties of the  model are peculiar.
In particular we found a Gaussian shape of the broadened
Landau levels and unusual expressions for the single particle
relaxation time $\tau\sb{s}$ and for the amplitude of the
de Haas--van Alphen oscillations.
These results were obtained by using the formalism of path
integrals in coordinate space which turns out to be the
most appropriate tool in this case.
In the present Letter, we apply these methods to the study of
magnetooscillations of the conductivity
in the FQHE regime near $\nu=1/2$.

We consider first the zero-temperature limit, when the fluctuations
of the fictitious magnetic field are determined by the randomly located
impurities \cite{hlr}.
We find the amplitude of the SDHO in this case to be proportional
to $\rho\sb{xx}\sp{osc}\propto
\exp [-(\pi /\omega\sb{c}\tau\sp{*}\sb{1/2})\sp{4}]$
where $\omega\sb{c}$ is the cyclotron frequency and
$\tau\sp{*}\sb{1/2}$
plays the role of an effective relaxation time.
This result which describes well the experimental data
should be contrasted with the usual behavior
$\rho\sb{xx}\sp{osc}\propto
\exp\left[-\pi /\omega\sb{c}\tau\right]$
that one obtains for the case of the short range random potential.

At finite temperature the particles are scattered additionally
by the thermal fluctuations of the magnetic field.
We find however that due to the screening in a strong
uniform magnetic field, these fluctuations give a contribution to the
single particle relaxation rate, which is small compared to
usual temperature-dependent factor $\exp(-2\pi\sp{2}T/\omega\sb{c})$
originating from the Fermi distribution.

We consider a realistic system formed by the 2D electron gas of density
$n\sb{e}$ and by the positively charged impurities located in a
layer separated by a large distance $d\sb{s}$ from the electron
plane. The statistical transformation attaches to each electron
an even number $\tilde{\phi}$ of flux quanta of the CS gauge field.
To describe the vicinity of the  $\nu=1/2$ state, we take
$\tilde{\phi}=2$; however the same formalism with
$\tilde{\phi}=4$ can be applied to the $\nu=1/4$ state.

In the mean field approximation, the statistical magnetic field
$B\sb{1/2}=4\pi c n\sb{e}/e$ cancels exactly the externally
applied field $B$ at $\nu=1/2$.
When the filling factor $\nu$ is tuned away from $\nu=1/2$,
the effective uniform magnetic field is equal to
$B\sb{e\!f\!f}=B-B\sb{1/2}$.
For  $\nu$ close to $1/2$, the number of filled Landau levels of
composite fermions
$p \gg 1$, so that the problem can be considered quasiclassically.
Essential components of the problem, in addition to the uniform
magnetic field $B\sb{e\!f\!f}$ are the random magnetic field $h(\bbox{r})$
and random potential $u(\bbox{r})$.
In the quasiclassical approximation, the quantities of interest
can be expressed as a sum over classical trajectories.
We will treat the random fields in the framework of quasiclassical
perturbation theory, neglecting their influence on the classical
trajectories. This approximation is valid provided
$\omega\sb{c}\tau\sb{t}\gg 1$.
The trajectories are then simply the cyclotron circles in the
uniform field $B\sb{e\!f\!f}$.

In a previous paper \cite{1} we used this quasiclassical approach
to calculate the de Haas--van Alphen oscillations of the density
of states (DOS) in the presence of a random magnetic field in the limit
$\omega\sb{c}\tau\sb{t}\gg 1$.
On this condition the conductivity can be written
as a sum over periodic orbits in a similar way \cite{abrikos,opp,rich};
the resulting expression has the same structure as for the DOS, except
for the overall prefactor representing the non-oscillating contribution.
We get therefore, in full analogy with \cite{1}
\begin{equation}
\sigma\sb{xx}=\sigma\sb{n\!o}\left[1-2i \mbox{Re}\sum\sb{k=1}\sp{\infty}\exp
\left\{2\pi ipk-{1\over 2}\left<S\sb{r}\sp{2}\right> k\sp{2}\right\}\right]
\label{sxx}
\end{equation}
where $\sigma\sb{n\!o}$ denotes the non-oscillating part of $\sigma\sb{xx}$,
summation goes over winding numbers $k$,
$p={1\over 2}{e\over c}R\sb{c}\sp{2}B\sb{e\!f\!f}
=2\pi {cn\sb{e}\over e B\sb{e\!f\!f}}$,
$R\sb{c}$ is the cyclotron
radius, $ S\sb{r}$ denotes the contribution
to the action induced by random fields
along the classical path of winding number $k=1$,
and the angular brackets denote the average over impurity configurations.
A good estimate for the amplitude of
the oscillations is given by the first harmonic in (\ref{sxx})
\begin{equation}
\sigma\sb{xx}=\sigma\sb{n\! o}
\left[1-2\cos\left(2\pi p\right)
\exp{\left(-{1\over 2}\left<S\sb{r}\sp{2}\right> \right)}\right] .
\label{sxx1}
\end{equation}

At zero temperature, the potential and magnetic fields fluctuations
are dominated by the randomly located impurities.
Each impurity creates a scalar potential of the form
\begin{equation}
\int(dq)v\sb{0}(q)e\sp{i\mbox{\boldmath$q$}(\mbox{\boldmath$r$}-
\mbox{\boldmath$r$}\sb{i})}\ ; \
v\sb{0}(q)={2\pi e\sp{2}\over \epsilon q}e\sp{-qd\sb{s}},
\label{coul}
\end{equation}
where $\bbox{r}\sb{i}$ is the projection of the impurity position
to the 2D plane,  $\epsilon$ is the dielectric constant,
and $(dq)=d\sp{2}q/(2\pi)\sp{2}$.
This potential gets renormalized due to the screening by fermions and
mixing with the CS field.
In the random phase approximation (RPA) one gets:
\begin{equation}
A\sb{\mu}=\left(\delta\sb{\mu}\sp{\
\rho}-U\sb{\mu\nu}K\sp{\nu\rho}\right)\sp{-1}
A\sb{\rho}\sp{(0)},
\label{dais}
\end{equation}
where we united scalar $A\sb{0}$ and vector $\bbox{A}$ potentials
in a covariant vector $A\sb{\mu}$; the vector $A\sb{\rho}\sp{(0)} $
represents the bare impurity potentials and therefore has only
$\rho=0$ non-zero component.
The tensors $U\sb{\mu\nu} $ and $K\sp{\nu\rho}$ represent the
bare gauge field propagator and the current-density response tensor
respectively.

To evaluate eq.(\ref{dais}) we use the Coulomb gauge $\mbox{div}\bbox{A}=0$,
go to the momentum space and choose the momentum $\bbox{q}$ to be
directed along the $x$-axis: $q\sb{x}=q$, $q\sb{y}=0$.
Then  $A\sb{\mu}$ has only 2 non-zero components corresponding to
$\mu=0$, $y$, and both K and U become $2\times 2$ matrices
\cite{hlr}:

\begin{eqnarray}
&&K\sp{\mu\nu}(q)=
\left(\begin{array}{cc}
-m\sp{*}/2\pi & -iq\sigma\sb{xy}\\ iq\sigma\sb{xy} & \chi
q\sp{2}-2i\omega n\sb{e}/qk\sb{F}
\end{array}\right) \nonumber \\
&&U\sb{\mu\nu}(q)=
\left(\begin{array}{cc}
v(q)  & 2\pi i \tilde{\phi}/q\\-2\pi i \tilde{\phi}/q  & 0
\end{array}\right) \label{tens} \\
&&A\sb{\mu}\sp{(0)}(q)=
\left(\begin{array}{c}
 v\sb{0}e\sp{-i\mbox{\boldmath$qr$}\sb{i}}  \\ 0
\end{array}\right) \ , \nonumber
\end{eqnarray}
where $m\sp{*}$ is the effective mass of fermions,
$\chi=1/12\pi m\sp{*}$ is the magnetic susceptibility and
$v(q)=2\pi e\sp{2}/(\epsilon q)$  is the Coulomb propagator,
and $\sigma\sb{xy}$ is the Hall conductivity of composite fermions
(not to be confused with electron  Hall conductivity).

Substituting (\ref{tens}) in (\ref{dais}), we find
\begin{equation}
A\sb{\mu}(q)=
{v\sb{0}(q)e\sp{-i\mbox{\boldmath$qr$}\sb{i}}\over
{ m\sp{*}v(q)\over 2\pi} + (\tilde{\phi}s+1)\sp{2}
+{\tilde{\phi}\sp{2}\over 6}}
\left(\begin{array}{c}
\tilde{\phi}s+1  \\ i\tilde{\phi} m\sp{*}/q
\end{array}\right),
\label{scr}
\end{equation}
where $s=2\pi\sigma\sb{xy}\simeq p$ in the limit
$\omega\sb{c}\tau\sb{t}\gg 1$ .

Let us now compare the first and the second term in denominator of
(\ref{scr}).
As we will see below, the typical momenta are $q\sim (2d\sb{s})\sp{-1}$,
and we get for $\tilde{\phi}=2$
\begin{equation}
{ m\sp{*}v(q)/2\pi\over (2s)\sp{2}}=
{m\sp{*}e\sp{2}\over 4\epsilon q s\sp{2}}\sim
{m\sp{*}e\sp{2}\over\epsilon k\sb{F}}{k\sb{F}d\sb{s}\over2p\sp{2}}\sim
{50\over p\sp{2}},
\label{est}
\end{equation}
where $k\sb{F}=\sqrt{4\pi n\sb{e}}$,
and we used  typical experimental parameters
\cite{st2} $n\sb{e}=1.1\cdot 10\sp{11}\mbox{cm}\sp{-2}$,
$d\sb{s}=80nm$, and the experimentally estimated value for the ratio
${m\sp{*}e\sp{2}/(\epsilon k\sb{F})}\sim 10$.
For the not too large $p$ we are interested in,
it is thus a reasonable approximation to neglect all but the first
term in the denominator of (\ref{scr}). This gives
\begin{equation}
A\sb{\mu}(q)=
{2\pi\over
 m\sp{*}} \tilde{\phi}e\sp{-i\mbox{\boldmath$qr$}\sb{i}}e\sp{-qd\sb{s}}
\left(\begin{array}{c}
p  \\ i m\sp{*}/q
\end{array}\right),
\label{scr1}
\end{equation}

The random field action $S\sb{r}$ in eqs.(\ref{sxx}), (\ref{sxx1})
is given by $S\sb{r}=-\oint{A\sb{\mu}dr\sp{\mu}}=
-\left(\int A\sb{0}dt -\oint \bbox{A}d\bbox{r}\right)$,
where the integration goes around a cyclotron orbit.
Assuming now the impurities to be randomly distributed with  concentration
$n\sb{i}$
and uncorrelated, we find
\begin{eqnarray}
&&\left<S\sb{r}\sp{2}\right>= (2\pi\tilde{\phi})\sp{2}n\sb{i}
\nonumber \\
&&
\times\int (dq)e\sp{-2qd\sb{s}}
\left|{p\over k\sb{F}}\oint dl\:
e\sp{-i\mbox{\boldmath$qr$}}+
\int d\sp{2}r\: e\sp{-i\mbox{\boldmath$qr$}} \right|\sp{2}
\label{act}  \\
&&=n\sb{i}(4\pi\sp{2}\tilde{\phi} R\sb{c})\sp{2}
\int (dq)e\sp{-2qd\sb{s}}
\left|{p\over k\sb{F}}J\sb{0}(qR\sb{c})+
{1\over q}J\sb{1}(qR\sb{c})\right|\sp{2}
\nonumber
\end{eqnarray}
Here $\oint dl$ means integration along the cyclotron orbit
and corresponds to the electric field contribution,
whereas $\int d\sp{2}r$
goes over the area surrounded by the orbit and describes the magnetic
field contribution.
Taking into account that $R\sb{c}\sp{2}=p\sp{2}/(\pi n\sb{e})$,
we have $R\sb{c}/2d\sb{s}=p/\sqrt{4\pi n\sb{e}d\sb{s}\sp{2}}\sim p/10
\lesssim 1$.
Thus for relevant momenta $q\sim1/(2d\sb{s})$ and  level numbers $p$,
 $qR\sb{c}\ll 1$ is a reasonable approximation.
In this case eq.(\ref{act}) reduces to
\begin{equation}
{1\over 2}\left<S\sb{r}\sp{2}\right>=
\pi\sp{3}\tilde{\phi}\sp{2}n\sb{i}
{R\sb{c}\sp{4}\over d\sb{s}\sp{2}}=
{n\sb{i}\over n\sb{e}}
{\pi\tilde{\phi}\sp{2}\over n\sb{e}d\sb{s}\sp{2}}
\left( {2\pi n\sb{e}\over m\sp{*}\omega\sb{c}}\right)\sp{4}.
\label{act2}
\end{equation}
Note that electric and magnetic field fluctuations give the same
contribution in this limit.
According to (\ref{sxx1}), this gives for the oscillating part
of the conductivity:
\begin{equation}
\sigma\sb{xx}\sp{osc}\propto -
\cos{\left({4\pi\sp{2}n\sb{e}c\over e B\sb{e\!f\!f}}\right)}
\exp{
\left[-\left({\pi\over\omega\sb{c}\tau\sb{1/2}\sp{*}}\right)\sp{4}\right]}
\label{sxxa}
\end{equation}
where we introduced a parameter $\tau\sb{1/2}\sp{*}$
which may be interpreted as an effective relaxation time and is
given according to eq.(\ref{act2}) by
\begin{equation}
\tau\sb{1/2}\sp{*}\simeq { m\sp{*}\over 2n\sb{e}}
\left({n\sb{e}d\sb{s}\sp{2}\over 4\pi}\right)
\sp{1/4}\ .
\label{tau}
\end{equation}
When writing eq.(\ref{tau}) we made the usual assumption that
concentrations of donors and charge carriers coincide:
$n\sb{e}=n\sb{i}$.

The dependence of the amplitude of oscillations on $\omega\sb{c}$
in eq.(\ref{sxxa}) differs from the conventional form
$\exp(-\pi/\omega\sb{c}\tau)$ which holds for short range
potential scattering.
We have already shown in \cite{1} that short range magnetic field
scattering leads to damping of oscillations in the DOS
$\sim\exp\left[-(\pi/\omega\sb{c}\tau)\sp{2}\right]$.
As we see now from eq.(\ref{sxxa}), a long range correlated magnetic
field leads to the novel result
$\sim\exp\left[-(\pi/\omega\sb{c}\tau)\sp{4}\right]$.

In Fig.1 we present experimental data for the amplitude
of $\rho\sp{osc}$ from \cite{st2}
($T=0.19K$, $B\sb{e\!f\!f}>0$).
It is seen that they can be fitted well by
$\exp\left[-(\pi/\omega\sb{c}\tau)\sp{4}\right]$,
whereas a $\exp(-\pi/\omega\sb{c}\tau)$ fit is much worse.

Now we consider the effect of dynamical fluctuations of the
gauge field at higher temperatures $T$.
At first glance, one could
expect those to be important for the following reason.
In the quasistatic approximation, the amplitude of magnetic field
fluctuations is given by \cite{ln}
\begin{equation}
\left< h(\bbox{r})h(\bbox{r}')\right>=\delta(\bbox{r}-\bbox{r}')T/\chi
\label{ht}\ .
\end{equation}
This would lead to a suppression of oscillations by a factor
$\sim\exp\left(-{1\over 2}\pi R\sb{c}\sp{2}T/\chi\right)\sim
\exp\left(-12\pi\sp{2}p T/\omega\sb{c}\right)
$
which could become  dominant  at high enough temperature.
However, eq.(\ref{ht}) was obtained in the absence of a uniform
magnetic field, whereas we are considering here the case of a relatively
strong field $B\sb{e\!f\!f}$.
As we will see below, this leads to a considerable weakening of magnetic
field fluctuations.

The propagator of gauge field fluctuations is given by
\begin{equation}
D\sb{\mu\nu}(q,\omega)=U\sb{\mu\rho}(q)
\left(\delta\sp{\rho}\sb{\ \nu}-
K\sp{\rho\lambda}(q,\omega)U\sb{\lambda\nu}(q)\right)\sp{-1}
\label{d}
\end{equation}
where $U\sb{\mu\nu}$ and $K\sb{\mu\nu}$ were given in eq.(\ref{tens}).
In particular for the $D\sb{11}$ component determining the magnetic
field fluctuations, we get
\begin{eqnarray}
&&D\sb{11}(q,\omega)=
\left(i\omega {2n\sb{e}\over
q k\sb{F}}-\tilde{\chi}q\sp{2}\right)\sp{-1} ;
 \nonumber \\
&&\tilde{\chi}=
{1\over2\pi m\sp{*}}\left[{1\over
6}+\left(s+{1\over\tilde{\phi}}\right)\sp{2}
\right]+{v(q)\over (2\pi\tilde{\phi})\sp{2}}.
\label{d11}
\end{eqnarray}
We see that due to the Hall conductivity $s=2\pi\sigma\sb{xy}$,
the value of $\tilde{\chi} $ exceeds considerably the bare value of
susceptibility $\chi=1/12\pi m\sp{*}$.
An analogous suppression of the gauge field fluctuations in the
external magnetic field was found by Ioffe and Wiegmann \cite{iw}
who considered the magnetoresistance of doped Mott insulators.

In the quasistatic approximation we find
\begin{equation}
\left< A\sb{1}A\sb{1}\right>\sb{q}=
\int{d\omega\over 2\pi}{2T\over\omega}\mbox{Im}D\sb{11}=
{T\over\tilde{\chi} q\sp{2}}=
{2\pi m\sp{*}\over s\sp{2}}
{T\over q\sp{2}}
\label{at}
\end{equation}
This leads to the following result for the amplitude of
magnetic field fluctuations:
\begin{equation}
\left< hh\right>\sb{q}={2\pi m\sp{*}\over s\sp{2}}T
\label{hta}
\end{equation}
The contribution of these fluctuations to the random field action
$\left< S\sb{r}\sp{2}\right>$ in (\ref{sxx1}) is therefore equal to
\begin{equation}
{1\over 2}\left< S\sb{r}\sp{2}\right>\sb{T}={1\over 2}
{\pi R\sb{c}\sp{2}}{2\pi m\sp{*}\over s\sp{2}}T=
{2\pi\sp{2}T\over p\omega\sb{c}}
\label{st}
\end{equation}

Thus the effect of dynamic fluctuations of magnetic field
to the amplitude of oscillations is small compared to the usual
Fermi factor
$(2\pi\sp{2}T/\omega\sb{c})/\sinh(2\pi\sp{2}T/\omega\sb{c})$
and can be neglected.

For completeness, we consider now the SDHO in low fields (around
$B=0$), where there is no CS gauge field.
At low temperature the scattering is then due to screened
impurity potential \cite{hlr}
$A\sb{0}=(2\pi/m\sb{b})e\sp{-qd\sb{s}}e\sp{-i\mbox{\boldmath$qr$}\sb{i}}$,
where $m\sb{b}$ is the band electron mass.
(It is easy to see that this follows from eq.(\ref{scr}) if one puts
$\tilde{\phi}=0$.)
We find that $\left< S\sb{r}\sp{2}\right>$ in eq.(\ref{sxx1}) is given for this
case
by
\begin{equation}
\left< S\sb{r}\sp{2}\right>
=(2\pi)\sp{2}{n\sb{i}\over n\sb{e}}R\sb{c}\sp{2}
\int qdq e\sp{-2qd\sb{s}}J\sb{0}\sp{2}(qR\sb{c})
\label{s0}
\end{equation}
Assuming again $n\sb{i}=n\sb{e}$ and $R\sb{c}\ll d\sb{s}$, we get for
the amplitude of oscillations
\begin{equation}
\sigma\sb{xx}\sp{osc}\propto -
\cos{\left({2\pi\sp{2}n\sb{e}c\over e B}\right)}
\exp{
\left[-\left({\pi\over\omega\sb{c}\tau\sb{0}\sp{*}}\right)\sp{2}\right]}\
,
\label{sxx0}
\end{equation}
\begin{equation}
\tau\sb{0}\sp{*}={ m\sb{b}\over n\sb{e}}
\left({n\sb{e}d\sb{s}\sp{2}\over \pi}\right)\sp{1/2}\ .
\label{tau0}
\end{equation}

As was already mentioned, the available experimental data around
$\nu=1/2$ \cite{st2} apparently show the behavior
$\ln \rho\sp{osc}\sim 1/\omega\sb{c}\sp{4}$
predicted by eq.(\ref{sxxa}).
The value of the parameter $\tau\sb{1/2}\sp{*}$
which is found from such a fit (Fig.2) is
$\tau\sb{1/2}\sp{*}\simeq 16\cdot 10\sp{-12}s$.
At the same time the theoretical estimate according to eq.(\ref{tau})
(with use of the parameters of \cite{st2}) gives
$\tau\sb{1/2}\sp{*}\simeq 2\cdot 10\sp{-12}s$
if one uses the experimental value of
the effective mass $m\sp{*}=0.7m\sb{e}$.
A similar discrepancy is found for the low--field relaxation time:
eq.(\ref{tau0}) for $m\sb{b}=0.07m\sb{e}$ gives
$\tau\sb{0}\sp{*}\simeq 1\cdot 10\sp{-12}s$,
whereas the value quoted in \cite{st2}
is $\tau\sb{0}\simeq 9\cdot 10\sp{-12} s$.
Comparison of these values should be taken
with certain caution, since  $\tau\sb{0}$ was found in \cite{st2}
by using a conventional linear fit for the Dingle plot,
rather than eq.(\ref{sxx0}).
We do not expect, however, that this could change  $\tau\sb{0}$
considerably.
We note also that the theoretically estimated values for the transport
relaxation rate at $\nu=1/2$ are typically $4$ times greater
than extracted from  experimental mobilities \cite{hlr,st2}.
Therefore the theory seems to overestimate relaxation
rates systematically.
 This situation is not new and has been discussed previously
\cite{col,lev}. The considerable increase of relaxation times can be
attributed to the correlations in positions of charged impurities
due to their mutual Coulomb interaction \cite{lev}.
It is not clear to us, however, whether the here obtained rather sizeable
discrepancy between experimental and theoretical values of
$\tau\sb{1/2}\sp{*}$ can be explained in this way or else implies an
inconsistency of the underlying theoretical picture
of $\nu=1/2$ FQHE.

In conclusion, we have presented a theory of magnetooscillations
around $\nu =1/2$ Landau level filling factor based on a model
with fluctuating Chern--Simons field \cite{hlr}.
The quasiclassical treatment of the problem is appropriate
and leads to unconventional
$\exp\left[-(\pi/\omega\sb{c}\tau\sp{*}\sb{1/2})\sp{4}\right]$
behavior of the amplitude of oscillations.
This result is in good agreement with available experimental data,
although the experimental value of $\tau\sb{1/2}\sp{*}$
exceeds our theoretical estimate.

This work was supported by German-Israel Foundation for Research
[GIF, Jerusalem, Israel] (E.A.), WE-Heraeus-Stiftung (A.G.A.),
the Alexander von Humboldt Foundation (A.D.M.)
and by SFB 195 der Deutschen Forschungsgemeinschaft (P.W.).

\begin{figure}
\caption{Dingle plot. Logarithm of normalized  amplitude of resistivity
oscillations $\ln({D\sb{T}\rho\sp{osc}/\rho})$, with \protect\\
$D\sb{T}=\sinh(2\pi\sp{2}T/\omega\sb{c})/(2\pi\sp{2}T/\omega\sb{c})$,
as a function of inverse effective magnetic field $B\sb{e\!f\!f}\sp{-1}$.
Linear, quadratic and quartic fits of experimental data from
\protect\cite{st2} are presented.}
\end{figure}
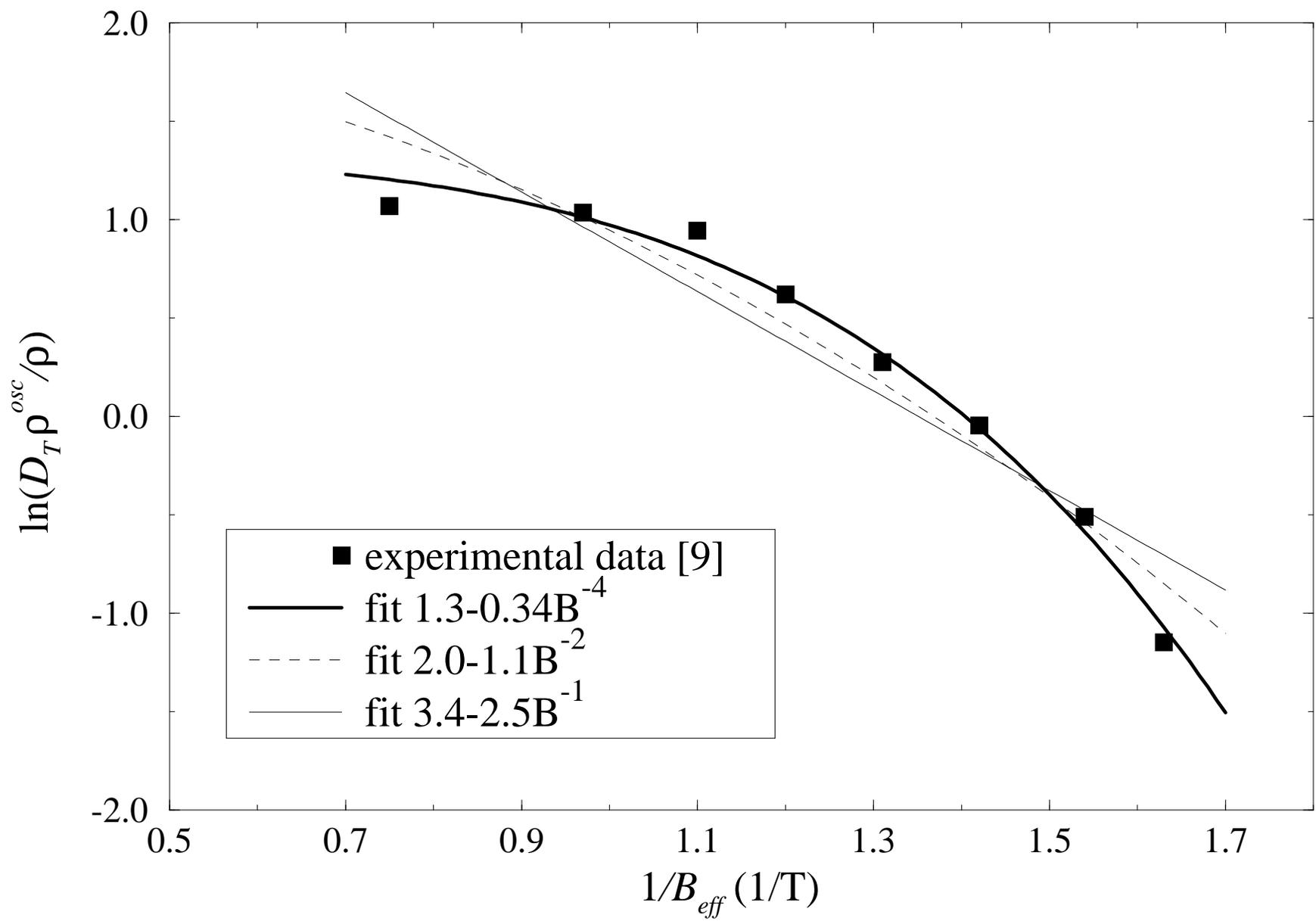
\end{document}